**A real time model to forecast 24 hours ahead, ozone peaks and exceedance levels.**

**Model based on artificial neural networks, neural classifier and weather predictions.**

**Application in an urban atmosphere in *Orléans*, *France*.**

Alain-Louis Dutot[a,]*, Joseph Rynkiewicz [b], Frédy E. Steiner [a], Julien Rude [a]

[a]Laboratoire Inter universitaire des Systèmes Atmosphériques-UMR-CNRS-7583

Université Paris12 et Université Paris7. 61 av. du Gal. De Gaulle 94010 CRETEIL Cedex, France.

[b]Laboratoire de Statistique Appliquée et Modélisation Stochastique. MATISSE-SAMOS-UMR-CNRS-8595

Université Paris 1 Centre Mendès France. 90 rue de Tolbiac 75634 Paris Cedex 13, France.

*Corresponding author.

Laboratoire Inter universitaire des Systèmes Atmosphériques-UMR-CNRS-7583

Université Paris12 et Université Paris7. 61 av. du Gal. De Gaulle 94010 CRETEIL Cedex, France. Tel. : 33 01 45 17 15 49 ; fax : 33 01 45 17 15 64.

*E-mail address*: dutot@lisa.univ-paris12.fr

**Abstract**

A neural network combined to a neural classifier is used in a real time forecasting of hourly maximum ozone in the centre of France, in an urban atmosphere. This neural model is based on the MLP structure. The inputs of the statistical network are model output statistics of the weather predictions from the French National Weather Service. These predicted meteorological parameters are very easily available through an air quality network. The lead time used in this forecasting is (t + 24) hours. Efforts are related: - to a regularisation method which is based on a BIC-like criterion – and to the determination of a confidence interval of forecasting. We offer a statistical validation between various statistical models and a deterministic chemistry-transport model. In this experiment, with the final neural network, the ozone peaks are fairly well predicted (in terms of global fit), with an Agreement Index = 92%, MAE = RMSE = 15 $\mu g\ m^{-3}$ and MBE = 5 $\mu g\ m^{-3}$, where the European threshold of the hourly ozone is 180 $\mu g\ m^{-3}$.

To improve the performance of this exceedance forecasting, instead of the previous model, we use a neural classifier with a sigmoid function in the output layer. The output of the network range from [0,1] and can be interpreted as the probability of exceedance of the threshold. This model is compared to a classical logistic regression. With this neural classifier, the Success Index of forecasting is 78% whereas it is from 65% to 72% with the classical MLPs. During the validation phase, in the Summer of 2003, 6 ozone peaks above the threshold were detected. They actually were 7.

Finally, the model called NEUROZONE, is now used in real time. New data will be introduced in the training data each year, at the end of September,. The network will be re-trained and new regression parameters estimated. So, one of the main difficulties in the training phase - namely the low frequency of ozone peaks above the threshold in this region - will be solved.

Software:

- **REGRESS**: Joseph Rynkiewicz (rynkiewi@univ-paris1.fr), Laboratoire de Statistique Appliquée et Modélisation Stochastique. MATISSE-SAMOS-UMR-CNRS-8595, Université Paris 1 Centre Mendès France. 90 rue de Tolbiac 75634 Paris Cedex 13, France. http://www.samos.univ-paris.fr. Running under LINUX and SOLARIS. Free of charge.

- **NEUROZONE**: Alain-Louis Dutot (dutot@lisa.univ-paris12.fr), Laboratoire Inter universitaire des Systèmes Atmosphériques-UMR-CNRS-7583, Université Paris12 et Université Paris7. 61 av. du Gal. De Gaulle 94010 CRETEIL Cedex, France. http://www.lisa.univ-paris12.fr. Visual Basic running under Windows. Free of charge. Software required : NETRAL : www.netral.com.

# 1) Introduction

According to the law, air quality agencies are commissioned to:

- monitor pollutants,
- forecast pollution peaks,
- inform authorities and public
- and assess the impact of emission reductions.

For all these reasons, there is nowadays a considerable challenge to air quality forecasting. Tools to forecast pollution peaks can be used in two different ways thanks to:

- 3-dimensional air quality models which integrate chemistry, transport and dispersion.
- or statistical models which generally directly connect meteorological conditions to level of pollutants.

The first approach is time-consuming and requires very large databases to initialize and run the model. Therefore, the second approach is generally preferred to the first one for real time forecasting. This study focuses on photochemical smog pollution and especially on ozone pollution. Although changes in daily emissions affect daily ozone concentrations, it is the daily weather variations that best explain the day to day variability in the ozone levels (US-EPA, 1999). Ozone conducive meteorological conditions are now well-known. During the Summer, high insulation, high temperature, high stability (low mixing heights), and low midday relative humidity produce photochemical smog. The persistence of these conditions leads to ozone episodes (Alshuller and Lefohn 1996, Seinfeld and Pandis 1998, US-EPA 1996). The method presented here, consists in using some of these meteorological parameters - estimated from a weather forecasting system - as the predictors of a statistical model.

# 2) Method

## 2-1) Strategy

The aim of this study is to present an ozone peak forecasting method which uses, in the statistical regression function, easily available variables for air quality agencies. In this study the predicted weather data will be provided by the model output statistics of the French National Weather Service (METEO-FRANCE).

Empirical ozone modelling and regression models in particular have been largely studied. The time series analysis can model the seasonality, the trend and the autocorrelation of the ozone variability. The regressive and autoregressive models are often used (Box and Jenkins 1976, Gonzalez-Manteiga et al 1993, Graf-Jacottet and Jaunin 1998, Prada-Sanchez et al 2000), but they are limited by the weakness of the modelling in the extreme values. Regressions are also associated with automatic classification as in the CART method (Gardner and Dorling 2000, Ryan 1995). The possible presence of chaotic dynamics in ozone concentrations allows the performing of non-linear time series modelling (Chen et al 1998, Kocak et al 2000, Lee et al 1994, Raga and Le Moyne1986). Other non-linear methods such as neural networks have also been developed (Boznar et al 1993, Comrie 1997, Gardner and Dorling 1998, Ruiz-Suarez et al 1995, Yi and Prybutok, 1996, Zolghadri et al, 2004). These neural methods provide a better representation of the extreme values than the linear ones. The use of non-linear techniques is often recommended to deal with the ozone prediction (Schlink et al, 2003, 2005). The aim of this paper is to test an empirical ozone modelling using a neural network coupled with a neural classifier to improve the performance of a threshold exceedance prediction.

### *2-2) Data set*

The ozone data used in this study come from LIG'AIR - the air quality agency of the centre of France. This agency has 15 ground monitoring stations distributed over 39 540 km$^2$. This region produces about 3% of the national NO$_x$ emissions and 10% of the VOC production, 38% of which come from biogenic emissions. In this region 85% of the NO$_x$ emissions come from mobile sources. 75% of the VOC regional emissions come from the industrial sector, waste processing and agriculture (http://www.citepa.org, http://www.ligair.fr). This study is focused only on the city of Orléans (274 000 inhabitants) which owns 3 urban monitoring stations: called **Préfecture** (station 1), **La Source** (station 2) and **Saint Jean de Braye** (station 3). The database for developing the model were obtained during 5 consecutive years (1999-2003). The ozone forecasting presented here is based on the daily maximum of the hourly mean concentrations of the background stations. As ozone peaks during the Summer, the data from April to September, are the only ones used in this model. This is the choice of the air quality network of Orléans, only to develop a "spring-summer" model. But, we have tested that our approach also is valid all over the year.

The meteorological data used are the weather forecasts delivered by the French Weather Service (*METEO-FRANCE-département du Loiret*) for the Orléans area. These forecasts are delivered at 12 00 UTC, for the following day. The forecast bulletin (D+1), called *atmogramme*, contains 5 classes of predicted parameters:

- the cloudiness, divided into 6 classes of weather conditions,
- the rainfall, divided into 10 classes,
- the surface wind speed and direction, divided into 6 and 8 classes respectively,
- the maximum and the minimum air temperatures,
- and the vertical temperature gradient between 0 and 300 m.

The first three parameters are predicted for 3-hour intervals and the last one for 12-hour intervals. We calculate the hourly frequency of each class of the 3 first types of parameters.

Persistence is introduced in the analysis by using ozone value at 12 00 UT, on D-day. These 74 meteorological variables and persistence are used as input data in the neural network. Note that the choice of these variables is directly imposed by the forecast bulletin and not by a statistical or chemical criterion. But it already contains the main part of the relevant meteorological parameters. Unfortunately, the solar radiation is not available. But it is highly correlated both to air temperature and cloud cover. Moreover, there are no data available on the upper-air ventilation reflecting the possible transport of the ozone and of the ozone precursors in and out of the site. The data used range from 1999 to 2003. Fig. 1 shows the variability of the ozone data.

### *2-3) Statistical model*

The neural model used in this study is autoregressive and includes exogenous parameters (it is termed a NARX model). It is based on the use of a MultiLayer Perceptron Network. From a statistical point of view, MultiLayer Perceptrons are non-linear parametric functions. The adjustable parameters are called the weight; the exogenous variables are the inputs of the model and the estimated variable is the output. One important feature of MultiLayer Perceptron is its capability to model any smooth functional relationship between one or more predictors and variables to be predicted. This property is completely fulfilled with just *one hidden layer MLP*. Thus, we have employed this kind of neural network, in this study. These regressions can be presented as static single-output processes with an n-input vector $X$ and an output vector $Y$. So, the estimated model, $\hat{Y}$, can be represented by:

$$\hat{Y} = f(X, w) + \varepsilon \qquad (1)$$

where: $f$ is the neural activation function

$w$ are the parameters of the neural regression to be estimated

and $\varepsilon$ is a zero-mean random variable.

Details on the use of such neural networks can be found, for example, in White (1992), Gardner and Dorling (1999), Nunnari et al (1998) and Dutot et al (2003). In this study, the hyperbolic tangent is used as the activation function, $f$, of the neural operator. So, equation (1) is written:

$$\hat{Y} = w_0 + \sum_{i=1}^{n} \left( w_i \tanh\left( w_{0,i} + \sum_{i=1}^{p} w_i X_i \right) \right) + \varepsilon \quad (2)$$

The two main pitfalls in using MLP are the poor local minima of the error function and the overtraining of this regression function. For the first problem, a good solution is to use a sufficiently large number of random initializations for the weights of the MLP to be trained. This method is time consuming but is easy to implement in parallel. It could also be convenient to use several computers to find the best parameters.

*2-3-1) Regularisation scheme*

Overtraining is a more complex problem. MLPs may be extremely overparametrized models. It occurs when the model learns the noisy details of the training data. The overtrained models have very poor performance on fresh data. In this study we have used a pruning technique to avoid overtraining. This technique, both for the parameter estimation (in the learning process) and the model selection (in the hidden layer architecture selection), is a stepwise method using a BIC-like criterion which has been proved consistent (Cottrell et al 1995). The MLP with the minimal dimension is found by the elimination of the irrelevant weights. The method essentially consists in minimizing a BIC-like information criterion, that is to say, the mean square error of the model penalized by a function of the number of parameters and data:

$$BIC = \ln \frac{MSE}{N} + W \frac{\ln(N)}{N}$$

where:

- $MSE = \frac{1}{N} \sum (X_{estimated} - X_{observed})^2$ = mean square error,

- $N$ = size of the training data

- and $W$ = number of adjustable parameters, $w_i$, of the MLP.

This minimization leads to the elimination of the irrelevant weights and, depending on the case, to the elimination of a few complete variables or even to the elimination of a few neural units in the hidden layer. This is a way to avoid overtraining of the learning phase. The determination of the final model begins with all the possible inputs and with too many neurons in the hidden layer. Then, units in the hidden layer are gradually eliminated by computing at each step the BIC-like criterion as long as its value decreases. The calculation stops when the *BIC* criterion remains stable or increases. In the beginning, the first value of the weights between the inputs and the hidden layer are initialized by the value of the parameters of a linear regression having the same inputs. The initial value of the bias is equal to the mean value of the training data. Finally, the elimination of these parameters, and eventually those of some inputs, leads to the minimal MLP. One advantage of this regularisation method instead of the classical early-stopping method, is to merge the learning and validation data sets into a bigger learning set. This methodology can be found in our software REGRESS, available at http//www.samos.univ-paris1.fr, running under LINUX and SOLARIS.

### *2-3-2) Confidence interval of prediction*

Monari and Dreyfus (2002) propose to use the leverage of the examples in the training data to compute a confidence interval of the predicted values. Leverage is a measure of the effect of a particular observation on the fitted regression, due to the position of the observation in the space of the predictor variables:

$$h_{ii} = Z_i^T M^{-1} Z_i$$

where:

- $h_{ii}$ is the leverage of the example $i$ in the training data which represents the influence of $i$ in the learning phase

- $Z_i = \dfrac{\partial f(x_i, \theta)}{\partial \theta}$ is the gradient of the model output with respect to the parameters $\theta$

- and $M = (Z^T Z)$.

The authors show that if the matrix $Z$ has full rank and under asymptotic conditions, then the confidence interval of prediction is:

$$\pm t_\alpha^{N-q} S \sqrt{\left( z^T \left( Z^T Z \right)^{-1} z \right)}$$

where:

- $t_\alpha^{N-q}$ is the t-distribution with $N-q$ degrees of freedom and a level of significance $(1-\alpha)$

- and $S = \sqrt{\left( \frac{1}{N-q} \sum_{i=1}^{N} R_i^2 \right)}$ is the residual standard deviation of the model.

All the details of this approach can be found on http://www.neurones.espci.fr.

In this work we have used a neural algorithm of the MLP developed by NETRAL (see http://www.netral.com). This software gives large access to the source code. The non-linear function used in the neural units is the hyperbolic tangent. The optimisation method used is a second-order method: the Levenber-Marquart method. The cost function to be minimized according to the Delta rule is:

$$E = \frac{1}{2} \Sigma \left( Y_{observed} - Y_{estimated} \right)^2.$$

Classically, the initial data are centered and standardized as:

$$X_{i,normalized} = \frac{(X_i - \overline{X})}{S_X}$$

where:

- $S_X$ is the standard deviation.

### 3) Results

*3-1) Inputs selection*

The neural network design will be obtained from the most representative station of the urban atmosphere: station 3, *Saint Jean de Bray*. This choice which includes location and environmental criteria is based on the national typology and classification of air quality monitoring sites. Then, the architecture (number of neurons in the hidden layer and number of input variables) of the network will be reproduced in the two other stations.

The database of the stations are split into 2 datasets, training and validation sets. The first dataset is used to optimise the parameters of the regression. The years 1999 to 2002 represent the training data, and 2003 the validation set. This validation dataset, which is not used during the training phase, is used to assess the performance of the regression. An ANOVA test indicated that there was not significant difference between these 2 datasets at the 95% confidence level.

According to the stepwise method (REGRESS) presented above, the variables selection leads to keep only 8 parameters as input data:

- the ozone concentration at 12 00 UTC on D-day,

- the predicted (D+1) minimum and maximum of the air temperatures,

- the predicted (D+1) mean surface wind speed,

- the predicted (D+1) hourly frequency of wind in directions: S, SW, W, NW and W,

- and the predicted (D+1) hourly frequency of a cloudiness class called "sky without cloud" in the *atmogramme*.

The REGRESS method also provides the number of neuronal units in the hidden layer. Fig 2 shows that the optimum of the BIC criterion, on the validation dataset, is reached with only 1 neuron in this hidden layer. Finally, Fig 3 presents the diagram of the neural network. This neural structure will be used on the other two stations, *Préfecture* (1) and *La Source (2)*.

*3-2) Details of model computations and model evaluation procedures*

The architecture of the basic neural model was shown above. We called this first model **MLP₁** in the rest of the study.

In Europe threshold for which an information of bad air quality in urban areas is made public is the hourly mean ozone concentration: 180 µg m$^{-3}$. Looking at Table 1, it is possible to see that the observations above the critical level of 180 µg m$^{-3}$ constitute a very small part of the overall data. Only 2% of the total data exceed this threshold. In case of such rare events, Nunnari et al (2004) propose a method of pattern balancing that is based on artificially reducing the frequency of data with low values. Let us call $N_a$ and $N_b$ the number of episodes above and below the threshold $\theta$. The authors propose to re-determine $N_b$ according to:

$$N^*_{b, MLP_2} = r(\theta) N_a$$

with $r(\theta) = a \exp(b\theta)$

where:

- $a=1$

- and $b=0.0125$ if $0 < \theta < 100$.

Though the European threshold is 180, we will implement this technique in the **MLP₂** model. As seen on Table 1, in this model the number of the training data are randomly reduced from 613 to 50 and 70 in the validation data.

According to the difference in the value of the threshold, a new balancing of the training set is proposed in this study. The size of the training set is empirically: $N^*_{b,MLP_3} = 2 N^*_{b, MLP_2}$. This model is called **MLP₃**.

A deterministic chemistry-transport model was then used to evaluate the other models. This model (called **CHIMERE**, Vautard et al, 2001) calculates, given the emissions, the meteorological variables and the lateral boundary conditions, the concentration fields of several pollutants, on a 6x6 km grid. The model produces the ozone forecasts for 4 different lead times: Day + 0, Day + 1, Day + 2 and Day + 3. The D+1 forecasts of this study will be used in comparison with the other models. Details on the model and the experiment can be found on our Web site: http://www.lisa.univ-paris12.fr and also on: http://prevair.ineris.fr.

Finally, we added the results of a multi-linear model (called **LIN**) which has the same predictors as the other models. To deal with obvious multicollinearity of the data we have used ridge regression. Instead of parameters estimation which is used in least square regression: $\hat{b}_i = (X^T X)^{-1}(X^T Y)$, the ridge regression use a biased estimator: $\tilde{b}_i = (X^T X + \lambda I)^{-1}(X^T Y)$, where I is the identity matrix. The ridge parameter $\lambda$ is the smallest value which gives stable estimate of $\tilde{b}_i$. In this study we have used $\lambda = 0.05$.

To evaluate the forecast system, pure persistence model is included in the performance indices. It will be called: **PERS**.

Generally, there are two main groups of performance measures that can be used in the evaluation: one group represents the global fit agreement between observed and predicted data and the other represents the quality of the forecasting exceedance of a threshold value. The first group contains:

- the Mean Bias Error (MBE) $= \frac{1}{N} \sum_{i=1}^{N}(P_i - O_i)$ which represents the degree of correspondence between the mean forecast ($P_i$ = predicted data) and the mean observation ($O_i$ = observed data). Values >0 indicate over-prediction.

- the Mean Absolute Error (MAE) $= \frac{1}{N} \sum_{i=1}^{N} |P_i - O_i|$

- the Root Mean Square Error (RMSE) $= \sqrt{\frac{1}{N} \sum_{i=1}^{N} (P_i - O_i)^2}$ which can be divided into systematic and unsystematic components by a least square linear regression of $P_i$ and $O_i$ with slope $b_1$ and intercept $b_0$. The systematic RMSE is RMSE$_s$ $= \sqrt{\frac{1}{N} \sum_{i=1}^{N} (\hat{P}_i - O_i)^2}$ where $\hat{P}_i$ is the predicted value in the linear regression: $\hat{P}_i = b_0 + b_1 O_i$. This measure describes the linear bias produced by the model. The unsystematic RMSE, RMSE$_u$ $= \sqrt{\frac{1}{N} \sum_{i=1}^{N} (\hat{P}_i - P_i)^2}$, may be interpreted as a measure of precision of the model. With respect to a coherent model, RMSE$_s$ should approach 0 while RMSE$_u$ should approache RMSE.

- the index of agreement, $d = 1 - \frac{\sum_{i=1}^{N}(P_i - O_i)^2}{\sum_{i=1}^{N}(|P_i - \overline{O}| + |O_i - \overline{O}|)^2}$ which should approach 1 in a coherent model.

- The correlation coefficient is not suitable for comparative study and therefore not used in this work (Willmot, 1985).

In the second group of indexes all observed and predicted exceedances are classified in a contingency table. With $A$ representing the correctly predicted exceedances, $F$ all the predicted exceedances, $M$ all the observed exceedances and $N$ the total number of data, we will use:

- the True Positive Rate, $TPR = A/M$ which represents the fraction of correctly predicted exceedances. It can be interpreted as the sensitivity of the model.

- the False Alarm Rate, $FAR = (F - A)/(N - M)$ for which $(1 - FAR)$ represents the specificity of the model.

- and the Success Index, $SI = TPR - FAR$ ranging from $-1$ to $+1$ with an optimal value of 1. Not affected by a large number of correctly forecasted non-exceedances, SI is useful for evaluating rare events as there are in this study.

Information about these performance indices can be found in Willmott et al (1985), European Environment Agency (1997) and Schlink et al (2003).

Note that the diurnal maximum 1-h average ozone reference level in Europe is 180 µg m$^{-3}$ (Directive 2002/3/EC).

### *3-3) Results of the statistical evaluation of the model performance*

Let us recall that the new data used to assess the models, come from Summer 2003 for station 3, the most representative one of the city.

#### *3-3-1) Evaluation of the global fit agreement*

Table 2 shows that the index of agreement, *d*, of the pure persistence model is the weakest of all the models which are all better than this reference model. But note that the 3 MLP obtain the highest scores for this general index (92%). Among these 3 models, MLP$_1$ is the most accurate (as RMSE unsystematic = 12 µg m$^{-3}$ and MAE = 15 µg m$^{-3}$). They are 17, 16, 26 µg m$^{-3}$ and 18, 17, 24 µg m$^{-3}$ for MLP$_2$, MLP$_3$ and CHIMERE respectively. Once again, the Mean Bias Error is lower, for MLP$_1$. Note that the 3 MLP models are slightly over-estimated (respectively: 5, 11 and 7 µg m$^{-3}$) whereas the deterministic model has a bias of $-13$ µg m$^{-3}$. The authors of this model explain the negative bias could be due to a problem in the representation of anthropic and biogenic emissions of VOCs during summer 2003. This problem now is solved. Finally, regarding these fit indices, we can select the MLP$_1$ model as an efficient forecasting system. Fig. 4 summarizes the results of Day+1 forecasts at station 3 during the validation period for both the $MLP_1$ and the deterministic models.

### 3-3-2) Evaluation of the exceedance indexes

The essential quality of a forecast model is its ability to correctly predict concentrations above the threshold. The success index (SI) can measure this ability. To evaluate the performance of the exceedance forecasting, we have calculated the frequency of the predictions such as:

a correct prediction is retained if $C_{estimated} + t_{\alpha}^{N-q} S \sqrt{\left( Z^T \left( Z^T Z \right)^{-1} Z \right)} \geq 180$

where $t_{\alpha}^{N-q} S \sqrt{\left( Z^T \left( Z^T Z \right)^{-1} Z \right)}$ is the confidence interval calculated as shown in paragraph 2-3.

For MLPs models, the best Success Index (72%) is obtained with the $MLP_3$ which uses our pattern balancing [see table 2]. But, this model also has one of the higher False Alarm Rate, 14%. This sensitivity gives 12 false alarms on the validation period while the observed exceedances only are 7. The model associated with the lowest FAR (6%) is the $MLP_1$ but this model also has about the same SI (65%) than the pure persistence model ! So, it is clearly difficult to do the right choice.

### 3-3-3) Development of a neural classifier to perform the exceedance levels

According to the difficulties to choose the right model of exceedance, we want to present in this work a significant improvement of the scores by using a new neural model. As seen above, the $MLP_1$ model is the best in terms of global fit. With the same neural structure, we have built a new neural network with a sigmoid neural function in the output layer instead of the identity function. The MLP network was defined as:

$$\hat{Y}_{MLP_1} = w_0 + \sum_{i=1}^{n} \left( w_i \tanh\left( w_{0,i} + \sum_{i=1}^{p} w_i X_i \right) \right).$$

The new neural network, which is called **classifier**, will be:

$$\hat{Y}_{classifier} = w_0 + \frac{1}{1+\exp\left[\sum_{i=1}^{n} \left(w_i \tanh\left(w_{0,i} + \sum_{i=1}^{p} w_i X_i\right)\right)\right]}.$$

All the input data are now ranging from $[0,1]$. The output of the network in the training data is no longer the ozone peak but $p_i$ with:

- $p_i = 0$ if $C_{estimated} + t_\alpha^{N-q} S\sqrt{\left(Z^T\left(Z^T Z\right)^{-1} Z\right)} < 180$

- $p_i = 1$ if $C_{estimated} + t_\alpha^{N-q} S\sqrt{\left(Z^T\left(Z^T Z\right)^{-1} Z\right)} \geq 180$

If:

- $N_A$ is the number of training data in the class $p_i = 1$

- $f_A(x)$ is the probability density function of the class $A$

- $N_B$ is the number of training data in the class $p_i = 0$

- $f_B(x)$ is the probability density function of the class $B$

then the function $P\langle A|x\rangle = \frac{N_A f_A(x)}{N_A f_A(x) + N_B f_B(x)}$ is the *a posteriori* class-conditional density function of $A$ given $x$.

Because of the sigmoid function in the output layer, the output o the classifier in the validation set, $\hat{Y}_{classifier, i}$, is also ranging from $[0,1]$. It can be interpreted as the probability of exceedance of the threshold 180 µg m$^{-3}$. This classifier is used on the same training data, with the same predictors and evaluated on the same validation set. If $\hat{Y}_{classifier, i} \geq 0.50$, we admit a prediction of exceedance. Table 2 shows that the SI of the classifier is the highest (78%) associated with one of the lowest FAR (8%). Using the best MLP model (MLP$_3$) these scores with a 95% confidence interval are 72% and 14% respectively. Fig. 5 shows the probability of exceedance during the validation phase. 6 ozone peaks have been detected above 180 µg m$^{-3}$ on 7 real alarms. And there are 6 false predictions. The correctly predicted exceedances for MLP$_3$ were only 4 out of 7. This method clearly shows a significant improvement in the prediction of the peak concentrations above 180 µg m$^{-3}$.

Is this method more accurate than a classical logistic regression ? A logistic model fits a response of observed proportions or probabilities at each level of an independent variable. The logistic model is defined by the equation:

$$y = \frac{\exp(a + \sum_i b_i x_i)}{1 + \exp(a + \sum_i b_i x_i)}$$

The response function of this model is a non-linear S-shaped curve with asymptotes at 0 and 1. In this study, after removing the no significant variables by using a stepwise technique, the best logistic model with the same data only uses 3 exogenous variables:

- the ozone concentration at 12 00 UTC on D-day,
- the predicted mean surface wind speed,
- and the predicted hourly frequency of wind in direction: SW.

And the equation of the logistic model is:

$$\text{probability of exceedance} = \frac{\exp(-12.23 - 0.8074\text{SW} - 0.4092\text{windspeed} + 0.091\text{ozoneonD} - \text{day})}{1 + \exp(-12.23 - 0.8074\text{SW} - 0.4092\text{windspeed} + 0.091\text{ozoneonD} - \text{day})}$$

The p-value of the model and the residual are 0.0001 and 1.00 respectively. Because the p-value of the model is less than 5%, there is a statistically significant relationship between the variables at the 95% or higher confidence level. In addition, the p-value of the residuals indicates that the model is not significantly worse than the best possible model for this data set at the 95% confidence interval. The Chi-square goodness of fit test p-value is 0.76. This determines whether the logistic function adequately fits the observed data. In this study there is no reason to reject the adequacy of the fitted model at the 95% or higher confidence level.

The results of this logistic regression during the validation phase are shown on Fig. 5. This figure shows that all the observed exceedances are correctly predicted. Especially the last isolated peak in the end of September, not seen by the classifier model, is correctly forecasted with the logistic model. Unfortunately the Fig. 5 also shows that this method is too much sensitive. There are 29 false alarms during this validation period. So, the Success Index of the logistic model becomes worse than those of the classifier: 66% and 78% respectively [see table 2].

Finally, the neural classifier remains the best candidate for the exceedance forecasting.

### *3-3-4) Forecasting on the other two stations*

The same neuronal structure ($MLP_1$), that is to say, the same predictors and the same hidden layer, is applied to the data of stations 1 and 2. A new training is made for each station with the data of 1999-2002 and a validation phase is made with the data of the year 2003. Fig.6 shows all the results for the 3 stations. The slopes of the linear regression of this scatter plot are $0.70 \pm 0.15$, $0.72 \pm 0.12$ and $0.70 \pm 0.15$ for station 1, 2 and 3 respectively. Let us remember that station 3 was used as reference. It is easy to see that there is no significant difference between these values. The analysis of the standardized residual, SR, ($SR = \frac{C_{estimated} - C_{observed}}{S}$ where $S$ is the standard deviation of the observed data) indicates that the variation of these values is less than $\pm 2S$, and that no special pattern appears in the residuals, Fig. 7. Therefore, these 3 models can be globally validated to evaluated the global fit agreement.

The neural classifier applied to the data of stations 1 and 2 gives the results shown on Table 3. All the observed exceedances are correctly forecasted for station 1. The SI has the higher score (89%). For the station 2, 6 exceedances out of 8 have been found. The SI is 70% for this station.

**4) Conclusion**

We have presented the results of an hourly maximum ozone forecast in an urban atmosphere. The system is based on an artificial neural network for which the input data come from meteorological model output statistics. Meteorological forecasts 24 hours ahead used in this model are easily available data provided by any National Weather Service. In this study, our aim was essentially to find these exogenous parameters and, of course, to develop an easily operational ozone system forecasting.

The validation procedure consists in comparing the forecasts with observations over a complete Summer season, from April to September 2003. One Multilayer Perceptron and two MLPs with a pattern balancing are tested. In comparison, we use a deterministic model and the persistence model as references.

For the global fit agreement, the MLP network without balancing shows the best scores of prediction 24 hours ahead with $d = 92\%$ and the best precision of the model with $MAE = 15\,\mu g\,m^{-3}$ and $RMSE_{unsystematic} = 12\,\mu g\,m^{-3}$.

For the ability to predict the exceedance of the European threshold (180 µg m$^{-3}$), the MLP with our balancing of the training data has the best Success Index (72%) but the False Alarm Rate is the higher (14%). According to the difficulty to choose the right model for exceedance forecasting, we have developed a neural classifier. The output of this model ranges from 0 to 1 and can be interpreted as the probability of the exceedance of the 180 threshold. Using this classifier during the validation phase, 6 exceedances out of 7 have been found. The Success Index reaches 78% and the False Alarm Rate only is 8%. This is the best combined performance of all the models.

Our model, now called *NEUROZONE*, has been implemented in real-time in the *Orléans* region. Each year, at the end of September, the validation data will be introduced into the former training data. Year after year, the network will be re-trained and new regression parameters estimated. So, one of the main difficulties in the training phase - namely the low frequency of ozone peaks above the threshold in this region - will be solved. In this study we have focused on a 24 hours ahead forecasting. In the future, it would be possible to introduce forecasting with longer lead times.

**Acknowledgements**

This forecasting project was supported by the air quality network of the Orléans region (**LIG'AIR**). We are also grateful to the French National Meteorological network (**METEO-FRANCE**) for the weather forecasts. **INERIS** graciously supplied us with the model-predicted data of the CHIMERE-PREV'AIR 2003 experiment, we gratefully acknowledge their assistance.

**Figures and Tables captions**

Fig. 1  Hourly maximum ozone from 1999 to 2003, during the Summer period - April to September.

Fig 2  BIC-like criterion according to the number of neurons in the hidden layer of the validation set. Note that the MLP with 0 neuron is a multilinear regression.

Fig. 3  Structure of the artificial neural network after the optimization by REGRESS.

Fig 4  Time series of the $MLP_1$ (in red) and the deterministic models (in green) [a] during the validation phase. [b] zooming on August 2003.

Fig 5  Probability of exceedance of the European threshold, during the validation phase (April to September 2003). Observations are in black, forecasted probability of the classifier are red squares and forecasted probability of the logistic model are blue squares.

Fig 6  Scatter plot of the measured and forecasted maximum ozone concentrations at the 3 ground-stations.

Fig 7  Standardized residual analysis on the validation dataset.

Table 1  Composition of the training and validation sets; in brackets the number of value >180 µg m$^{-3}$, outside the total number of data.

Table 2  Performance measures of all the models during the validation phase: April to September 2003.

Table 3  Performance measures of the exceedance forecasting for the 3 stations.

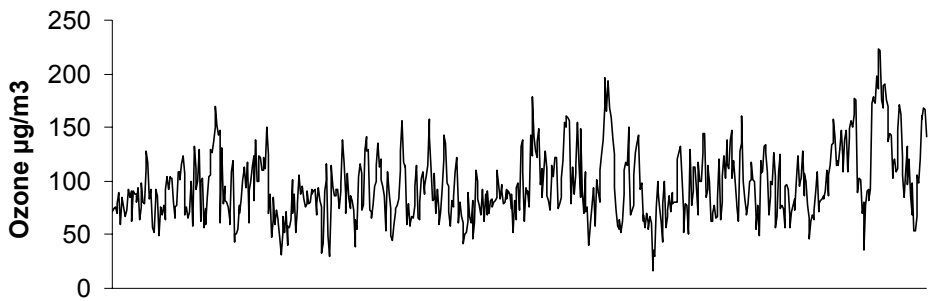

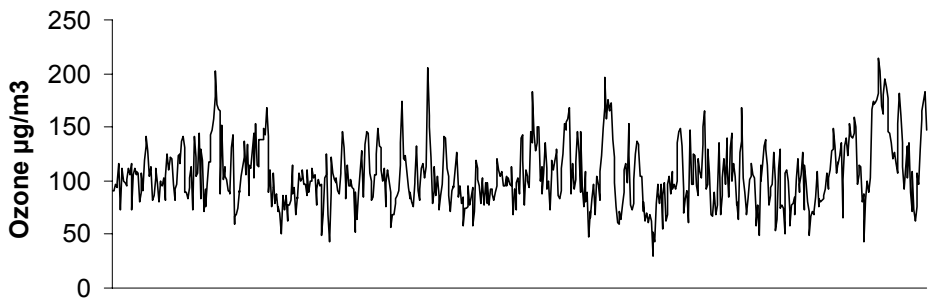

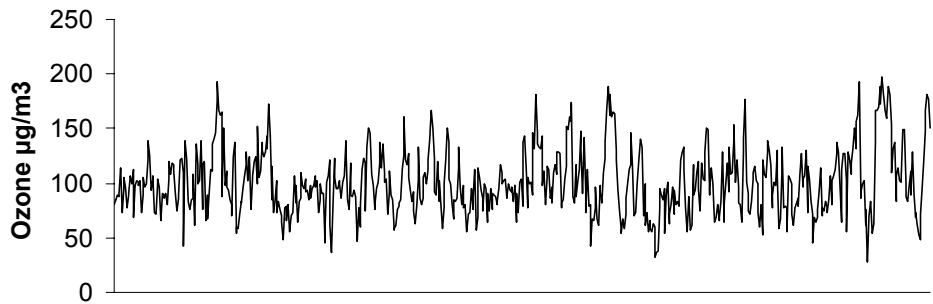

Fig. 1  Hourly maximum ozone from 1999 to 2003, during the Summer period - April to September.

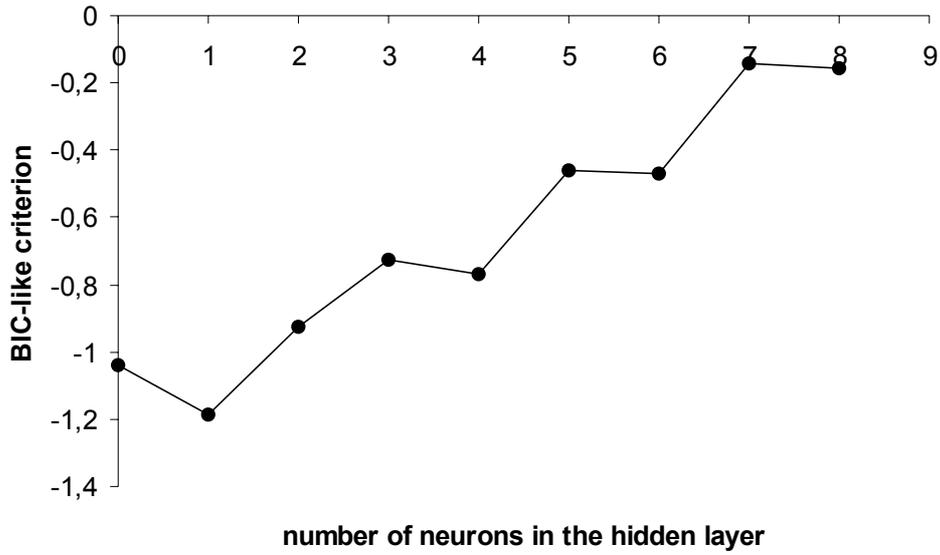

Fig 2 BIC-like criterion according to the number of neurons in the hidden layer in the validation phase. Note that the MLP with 0 neuron is a multi-linear regression.

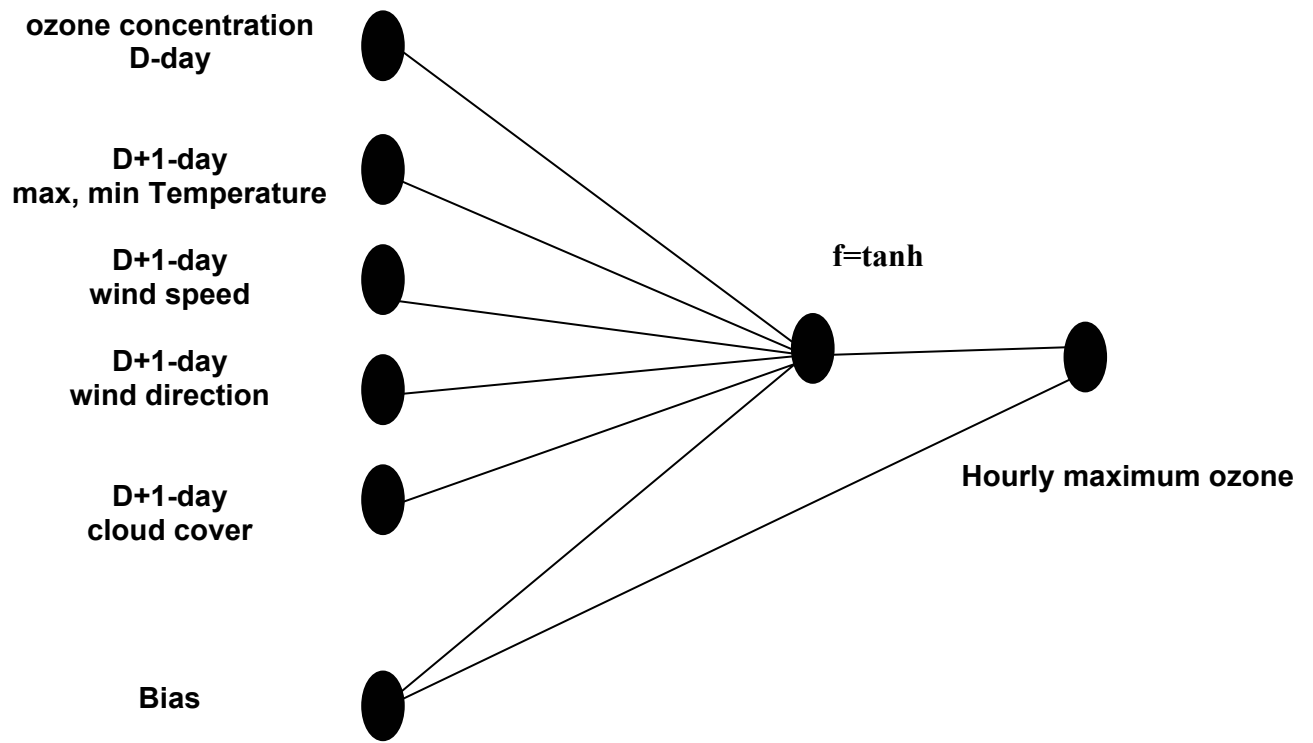

Fig. 3 Structure of the artificial neural network after the optimization by REGRESS.

**(a)**

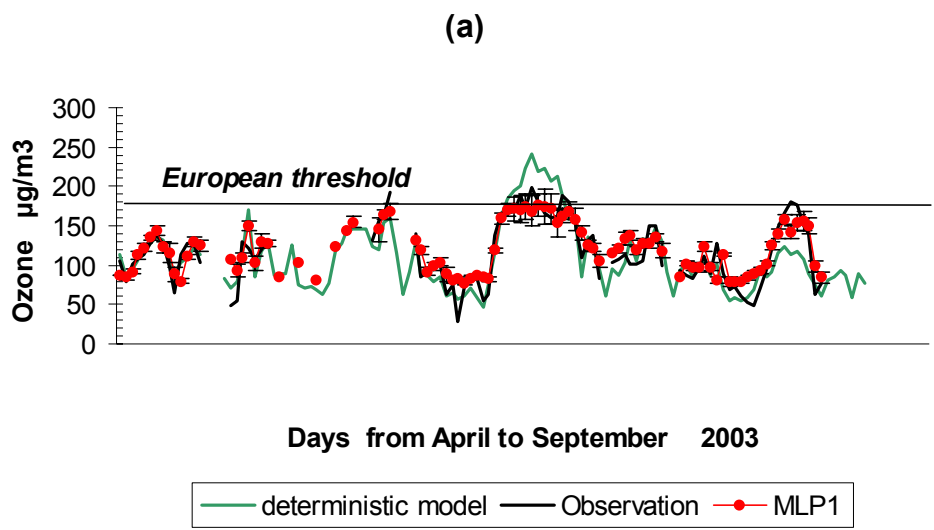

**(b)**

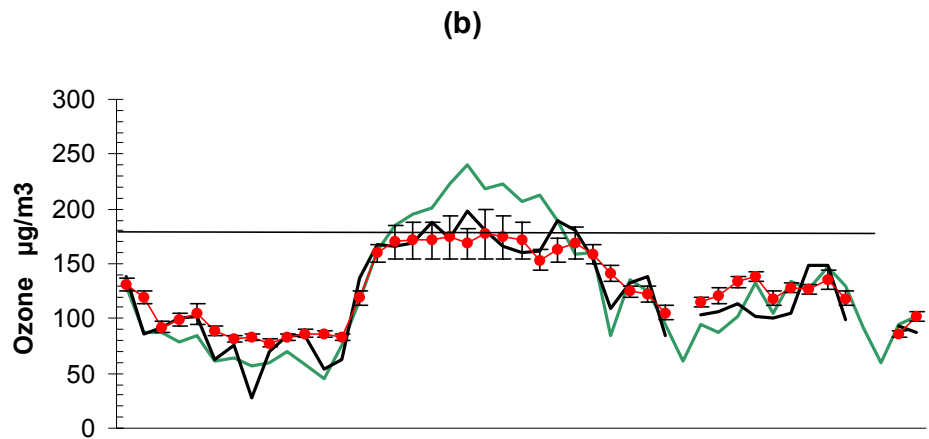

Fig 4  Time series of the MLP$_1$ (in red) and the deterministic models (in green) [a] during the validation phase. [b] zooming on August 2003.

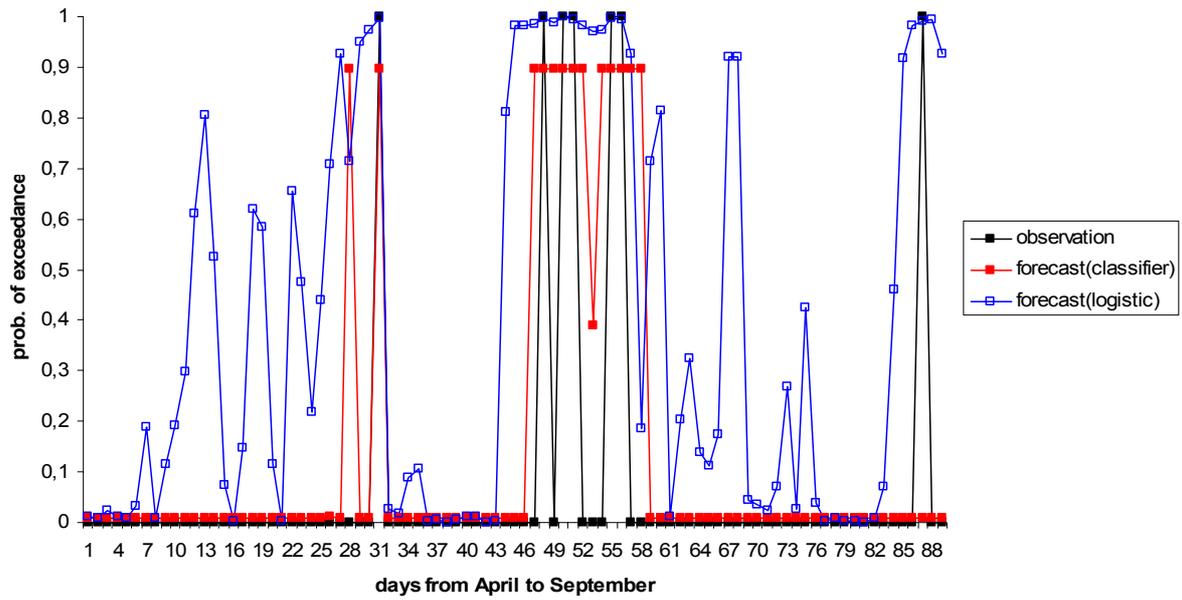

Fig 5 Probability of exceedance of the European threshold, during the validation phase (April to September 2003). Observations are in black, forecasted probability of the classifier are red squares and forecasted probability of the logistic model are blue squares.

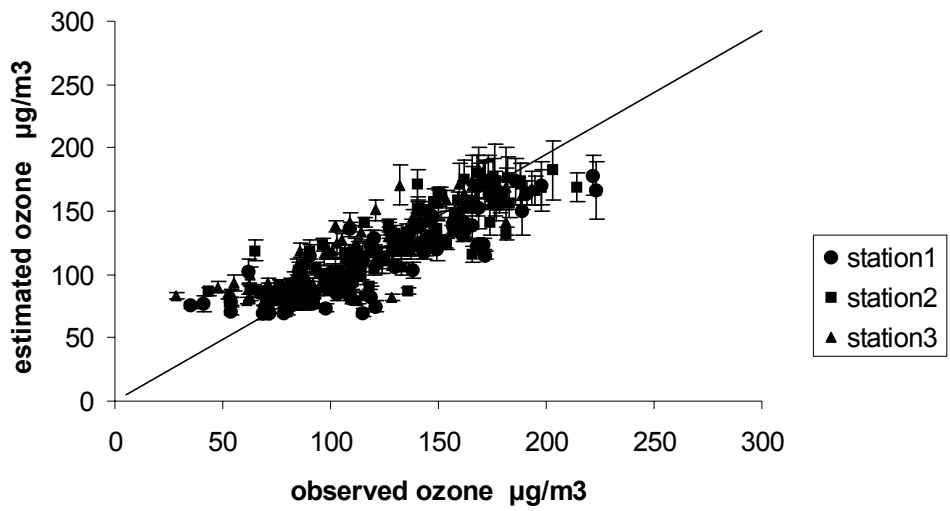

Fig 6 Scatter plot of the measured and forecasted maximum ozone concentrations at the 3 ground-stations.

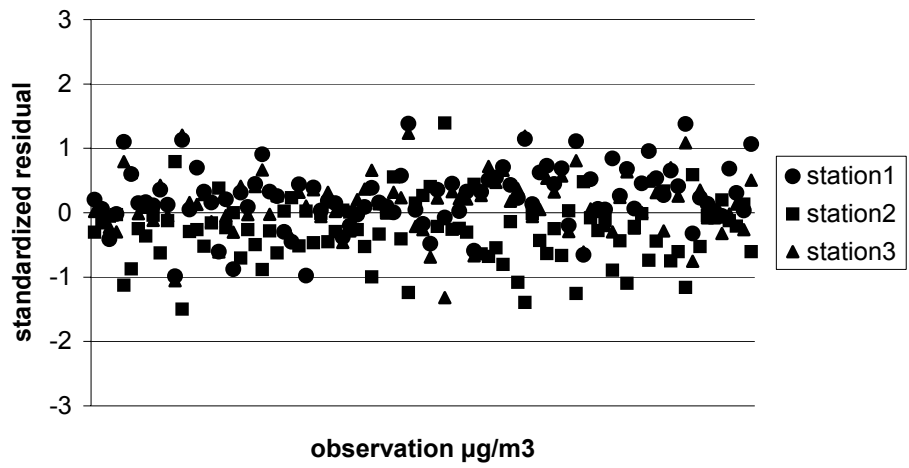

Fig 7  Standardized residual analysis on the validation dataset.

|  | **MLP$_1$** | **MLP$_2$** | **MLP$_3$** |
|---|---|---|---|
| **1999 training** | 159(1) | 10(1) | 20(1) |
| **2000 training** | 161(0) | 0(0) | 0(0) |
| **2001 training** | 162(3) | 30(3) | 60(3) |
| **2002 training** | 131(1) | 10(1) | 20(1) |
| **total training** | 613(5) | 50(5) | 100(5) |
| **2003 validation** | 105(7) | 70(7) | 105(7) |

Table 1. Composition of the training and validation sets; in brackets the number of value >180 µg m$^{-3}$, outside the total number of data.

|            | MBE | MAE | RMSE | RMSE$_s$ | RMSE$_u$ | d    | FAR  | SI   |
|------------|-----|-----|------|----------|----------|------|------|------|
| **PERS**   | -2  | 20  | 20   | 8        | 19       | 0,88 | 0.07 | 0,64 |
| **LIN**    | 5   | 17  | 14   | 14       | 11       | 0,90 | 0.02 | 0.12 |
| **MLP$_1$** | **5** | **15** | **15** | 11   | **12**   | **0,92** | 0.06 | 0.65 |
| **MLP$_2$** | 11  | 18  | 18   | 12       | 17       | **0,92** | 0.17 | 0.69 |
| **MLP$_3$** | 7   | 17  | 18   | 8        | 16       | **0,92** | **0.14** | **0.72** |
| **CHIMERE** | -13 | 24  | 30   | 14       | 26       | 0,89 | 0.05 | 0.52 |
| **Classifier** | - | - | -  | -        | -        | -    | **0.08** | **0.78** |
| **Logistic** | - | -  | -    | -        | -        | -    | 0.34 | 0.66 |

Table 2 Performance measures of all the models during the validation phase: April to September 2003.

|  | Station 1 | Station 2 | Station 3 (reference) |
|---|---|---|---|
| **FAR** | 11% | 5% | 8% |
| **SI** | 89% | 70% | 78% |
| **Observed peak** | 6 | 8 | 7 |
| **Forecasted peak** | 6 | 6 | 6 |
| **False alarm** | 10 | 4 | 6 |

Table 3  Performance measures of the exceedance forecasting for the 3 stations.